\documentstyle[11pt,aaspp4]{article} 
\slugcomment{{\it submitted to The Astrophysical Journal Letters 
	on January 7, 1999}}

\begin{document}

\title{Objects in NGC~205 Resolved into Stellar Associations by HST
    Ultraviolet Imaging\footnotemark}

\footnotetext{Based on observations with the NASA/ESA Hubble Space
    Telescope, obtained at the Space Telescope Science Institute, which
    is operated by AURA, Inc., under NASA Contract NAS 5-26555.}

\author{Michele~Cappellari\altaffilmark{2},
    Francesco~Bertola\altaffilmark{2}, David~Burstein\altaffilmark{3},
    Lucio~M.~Buson\altaffilmark{4}, Laura~Greggio\altaffilmark{5,6}\ \&
    Alvio~Renzini\altaffilmark{7}}

\altaffiltext{2}{Dipartimento di Astronomia,
    Universit\`a di Padova, Padova, Italy}
\altaffiltext{3}{Department of Physics \& Astronomy,
    Arizona State University, USA}
\altaffiltext{4}{Osservatorio di Capodimonte, Napoli, Italy}
\altaffiltext{5}{Osservatorio di Bologna, Bologna, Italy}
\altaffiltext{6}{Universitaets Sternwarte, Muenchen, Germany}
\altaffiltext{7}{European Southern Observatory,
    Garching bei M\"unchen, Germany}

\begin{abstract}

We have obtained high resolution UV images with the HST+WFPC2 of the
central region of the dwarf elliptical galaxy NGC~205. Our images reveal
that many of the hot UV stars previously detected and studied from the
ground are actually multiple systems, open clusters and star associations.
We have performed photometry of two such clusters and we find our data are
consistent with stellar ages of 50 and 100 Myr, respectively. From the
number of massive stars in NGC 205 we estimate that the star formation
episode in this galaxy has turned $\sim 1,000\, M_\odot$ of gas into stars
over the last 100 Myr.

\end{abstract}


\section{INTRODUCTION}

Since the first observations by Hubble (1932) NGC~205, one of the dwarf
elliptical companions to M31, has been known to contain a sprinkling of
young stars.  This is evidenced by ground-based resolved OB associations
(Hodge 1973), dust obscuration (Hubble 1932; Hodge 1973; Peletier 1993; Lee
1996) and presence of molecular hydrogen (Young \& Lo 1996), as well as
several early-type stars with unusually blue \ub\ colors (e.g.,
$\ub\le-1.0$; objects labeled 4, 7 and 12 in the study of Peletier 1993).
In our post-COSTAR program to image the hot stars among old stellar
populations in Local Group galaxies, we obtained WFPC2 UV images of
NGC~205.  We find that many of the brightest ``stars'' in NGC~205 are
actually multiple systems, including {\it bona fide} examples of
Orion/Pleiades-type stellar systems.

Indeed, many OB and early-type associations near the Sun have typical sizes
of $\sim2-5$ pc, including the Orion OB association ($\sim1^\circ$ at 460
pc) and the Pleiades ($\sim2^\circ$ at 116 pc). At the distance of NGC~205
(830 kpc, Lee 1996) these associations would subtend an angle $\sim1''$,
difficult to resolve from the ground.  Another classical example is offered
by star-forming regions within the Large Magellanic Cloud (e.g., R136 in 30
Dor, cf. Weigelt et al.\ 1991), where objects previously thought to consist
of a few stars from ground-based observations resolve themselves into
hundreds, if not thousands, of stars at the angular resolutions of HST,
i.e., 0\farcs05.

In this letter we present the pictorial evidence for the existence of these
kinds of stellar groupings in NGC~205, which were previously considered to
be single stars.  The observations are presented in Section~2, while in
Section~3 the photometry of individual objects in two ultraviolet bands is
presented and the derived color-magnitude diagrams are compared to
theoretical isochrones. Our main conclusions are summarized in Section~4.

\section{OBSERVATIONS}
 
HST/WFPC2 observations of NGC~205 (GO 5430) were taken on 30 September 1994
through the WFPC2 filters F185W (9600 s), F255W (9000 s), F336W (1200 s)
and F675W (150 s). These images are calibrated using the standard STScI
pipeline with the latest available WFPC2 calibration files, then coadded
and cosmic-ray-cleaned using the IRAF/STSDAS task CRREJ.
Figure~\ref{mosaic_stars} presents a mosaic of 2\farcs8$\times$2\farcs8
regions centered on each of the 10 sources identified as stars on
ground-based images by Peletier (1993; using his numbering system), which
also are in our WFPC2 frames. Figure~\ref{mosaic_archipelago} shows a
similar mosaic for a larger region 20$''$ north of the nucleus rich in
UV-bright stars (Hodge 1973; Peletier 1993; Lee 1996), which we term the
``Archipelago''. The nucleus of NGC~205 is also shown for reference.

Objects 2, 3, 4, 5, 6 and the nucleus (``C'') fall in the PC1 CCD of the
WFPC2 (pixel size 0\farcs046).  The other objects are in the WF2 CCDs
(pixel size 0\farcs100). As the WF CCDs heavily undersample the
$\sim0\farcs05$ FWHM PSF, we remove the pixellated appearance of the data
by bilinear interpolation onto a finer grid with pixel size 0\farcs100/2.

As is evident, at least half of the brightest ``stars'' identified by
Peletier (1993) are actually multiple systems of UV-bright sources. Objects
1 and 8 are likely open clusters, as indicated by their size ($\sim 2$ pc,
similar to that of the Orion and Pleiades clusters) and numbers of bright,
resolved stars. Objects 3 and 4 are multiple systems. Object 11 is likely a
background galaxy, judging from its diffuse appearance in the F675W image
and its faintness in the UV images. Most of the UV-bright stars in the
Archipelago are multiple on scale sizes less than $1''$. The two clusters 1
and 8 shown in Figure~\ref{mosaic_stars}, as well as the Archipelago
(Figure~\ref{mosaic_archipelago}) are identified in the F336W filter image
(Figure~\ref{all_field}).

Previous HST observations have shown that the nucleus of NGC~205 consists
of a globular-cluster-sized stellar group (cf. Bertola et al.\ 1995, Jones
et al.\ 1996). We clearly resolve the nucleus of NGC~205. Its axial ratio
is $q=0.90\pm0.05$ (there is some evidence that it is more elongated in the
UV). Its position angle is $P.A.=166\arcdeg\pm8\arcdeg$, consistent within
the errors with previous measurements by Bertola et al.\ (1995) and Jones
et al.\ (1996), as well as with the position angle for the whole galaxy
(Nilson 1973). The centroid of the nucleus with respect to the stars is the
same in all HST images.

\section{PHOTOMETRY}

We used the DAOPHOT PSF-fitting package within the IDL Astronomical Library
(Landsman 1995) to photometer individual stars in the Clusters 1 and 8 and
in the Archipelago.  A $4\sigma$ detection limit is adopted, equivalent to
limiting magnitudes $m_{336}\simeq22.9$ and $m_{255}\simeq22.7$. Owing to
the lack of good S/N observations of isolated stars in our field, we have
fitted a synthetic PSF generated by the Tiny Tim software (Krist 1993),
specifically for each different band.

We find 24 UV-bright stars in the Archipelago, 7 in Cluster 8 and 4 in
Cluster 1.  These data are used to create the color-magnitude diagram of
Figure~\ref{isocrone_n205}. Adopted magnitudes are in the STMAG system
($m_{ST}=-21.10-2.5\log f_\lambda$, with $f_\lambda$ in erg cm$^{-2}$
s$^{-1}$ \AA$^{-1}$). The effect of crowding and the severe undersampling
of the PSF does not allow a reliable measurement of the other fainter
cluster members otherwise visible in both the F336W and the F255W resampled
images of Figure~\ref{mosaic_stars} and Figure~\ref{mosaic_archipelago}.

Figure~\ref{isocrone_n205} also shows Bertelli et al.\ (1994) solar
metallicity theoretical isochrones for ages of 20, 50 and 100 Myr,
corrected for a total reddening of $E(\bv)=0.18$ and a
foreground-extinction-corrected distance modulus of $m-M=24.6$ (Lee 1996).
The theoretical isochrones are converted to the observational plane using
the model atmospheres of Kurucz (1979a,b). The model spectra have first
been reddened using the Cardelli et al.\ (1989) extinction curve (for
$A_{\rm V}/E(\bv)=3.2$). Convolution of the reddened spectra with the most
up-to-date instruments+filters sensitivity in the SYNPHOT package within
IRAF/STSDAS then yields magnitudes and colors.

Our estimate of $E(\bv)$ has been determined directly from our data. A
lower reddening would make the isochrones too blue relative to the observed
stars. This value of the reddening is identical to that determined by Han
et al. (1997) for the foreground reddening of NGC 147, another companion of
M31, but is significantly larger than the value $E(\bv)=0.035$ given by
Burstein \& Heiles (1982). For the origin of this discrepancy it is
difficult to disentangle between an excess Galactic extinction with respect
to the Burstein \& Heiles map, and internal absorption to NGC 205 itself.

The combination of low angular resolution, stellar crowding in clusters and
relatively shallow limiting magnitude limits the stars in our CM diagram
(CMD) to the brighter end of the main sequence (MS), and a few early
post-MS stars. Crowding affects our CMD in two ways: an artificial
brightening of the MS stars, and an artificial reddening of the measured
star relative to the theoretical MS. This latter effect is analogous to
what happens in an unresolved double star.

As shown in Figure~\ref{isocrone_n205}, the adopted mean reddening,
distance modulus, and isochrone metallicity give a reasonable match to the
distribution of stars in the CMD.  The scatter seen is consistent with
observational error.  The Archipelago region contains stars with a range of
age, with some as young as 20 Myr and masses up to $\sim 11\, M_\odot$ and
ones as old as 100 Myr and masses of $\sim 5\, M_\odot$. These objects are
expected to spend their core helium burning phase at colors redder than
$m_{255}-m_{336}\simeq2$, i.e., too red with respect to the color of the
two reddest stars in the Archipelago. These two stars are likely blends of
blue main sequence and red core helium burning stars. The four brightest
stars in Cluster 1 are likely main sequence stars as young as the youngest
stars in the Archipelago region.  In contrast, the stars in Cluster 8 have
UV brightness and colors located between the main sequence turn-offs of the
50 and 100 Myr isochrones.  While estimates of age ($\sim 70$ Myr) and mass
($6\, M_\odot$) can be made from Figure~\ref{isocrone_n205} for the stars
in Cluster 8, uncertainties owing to internal dust and photometric error
make these estimates rather uncertain.  What is apparent is that the stars
in Cluster 8 are older by $\sim 50$ Myr than those in Cluster 1.

\section{DISCUSSION}

It is not surprising that objects that on poor resolution images appear as
single stars turn out to be resolved into many stars with the far superior
angular resolution of HST. Warnings on the risks of overlooking this effect
of crowding date from the 1950's and 1960's, when Hubble and Sandage (see
Sandage \& Hubble 1961) noticed the problem for the bright red supergiants
in the Local Group. More recent examples include the HST observations of
the stellar population of M32 (Grillmair et al.\ 1996), and the classical
case of R136 resolved into myriads of stars (Weigelt et al.\ 1991).  The
effect of stellar blends are discussed in general terms by Renzini (1998),
with several examples drawn from stellar populations in Local Group
galaxies.

Local Group galaxies are currently the only ones for which we can
reconstruct their star formation history in a fairly direct and detailed
way from the photometry of individual stars. Crowding, however, is a major
problem. Our results demonstrate that we have to be wary of interpreting
bright ``stars'' seen with insufficient resolution as individual stars,
even in the case of what appear to be bright, young stars in an otherwise
diffuse dwarf galaxy.  As is apparent from the images presented in this
paper, it is quite common that the brightest objects are actually not
individual stars.  This is to emphasize once more that the reconstruction
of the star formation history of Local Group galaxies requires superior
angular resolution imaging. Blends produce brighter objects, which if
interpreted as single stars lead to infer younger ages than the real ones,
no matter whether dealing with young main sequence stars, very old red
giant branch, or asymptotic giant branch stars.

In the case of NGC~205, young objects are confined to within $\sim1'$ from
the center (Peletier 1993), and are virtually all included in our WFPC2
frames. On these images we can count a total of $\sim35$ stars likely to be
in the mass range between $\sim5$ and $\sim10\,M_\odot$.  Based on their
positions in the UV color-magnitude diagram, and assuming a normal IMF, the
star formation episode that produced the Archipelago and the young star
clusters in NGC~205 has turned $\sim1,000\,M_\odot$ of gas into stars
within the past 100 Myr.  Therefore, we are dealing with a quite modest
star formation event, of order of $10^{-5}\,M_\odot$/yr when averaged over
the last $10^8$ years. Even with HST, an event of this kind would become
unobservable in a few $10^8$ years, and cannot contribute much to the
stellar mass of NGC~205. The question as to where this gas came from is not
easy to answer. Possibly, as in our Galaxy, high velocity HI clouds may be
orbiting M31, and one such cloud might have been captured by the potential
well of NGC~205 and squeezed above star formation threshold.

\acknowledgments

DB acknowledges support by NASA/STscI through grants GO-03728.01-91A and
GO-06309.01-94A to DB. LG is grateful to the Observatory of the University
of Munich for its extensive hospitality and for the providing a most
stimulating environment.


{}



\clearpage
\epsscale{.4}
\plotone{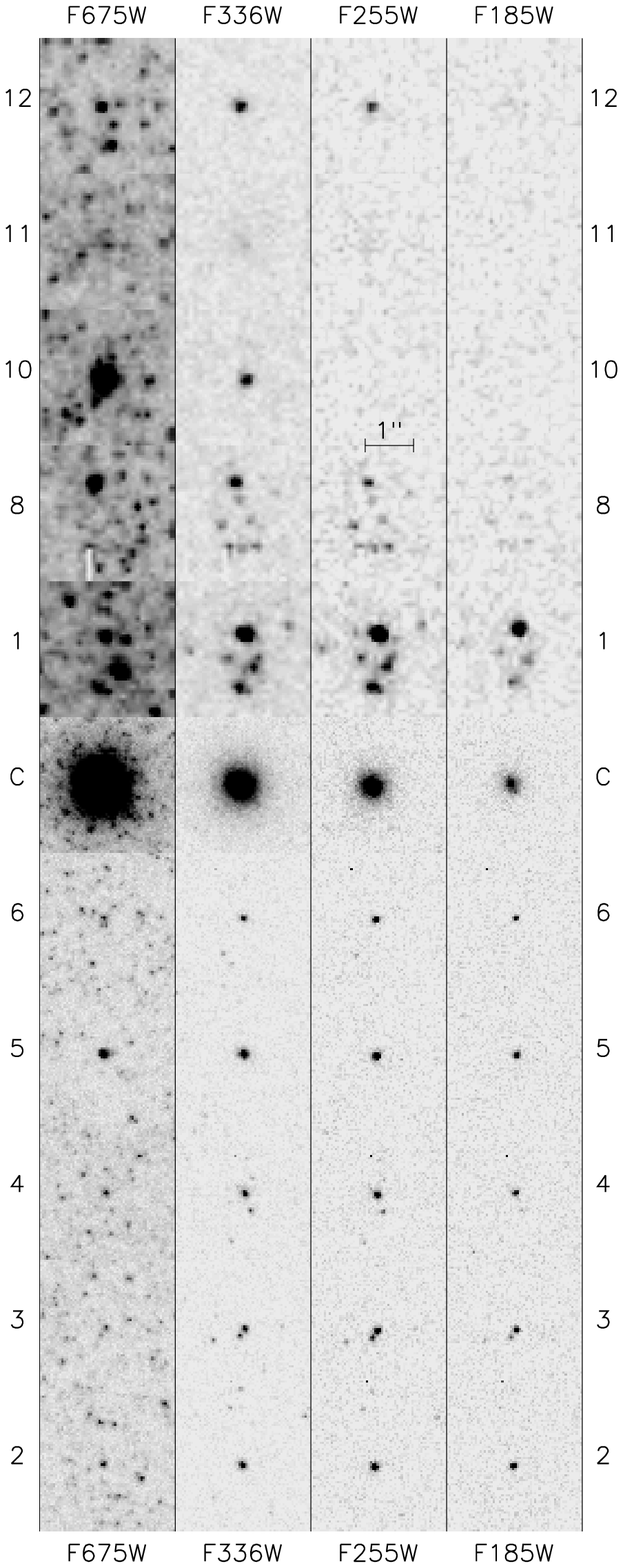}
\figcaption{The WFPC2 F675W, F336W, F255W and F185W
2\farcs8$\times$2\farcs8 FOV images (from left to right, respectively) of
ten ``stars'' listed by Peletier (1993) in the central region of NGC~205,
and of the galaxy nucleus (labeled as ``C''). Objects labeled 2--6 and the
nucleus ``C'' were included in the PC1 CCD which has a pixel size of
0\farcs046.  Frames 1, 8, 10, 11, and 12 are taken from the WF2 CCD which
has a pixel size of 0\farcs100, and have been resampled into a finer grid
with pixel size 0\farcs100/2 by means of bilinear interpolation to smooth
the pixellated appearance of the image. Orientation as in
Figure~\ref{all_field}.
\label{mosaic_stars}}

\clearpage
\epsscale{.28}
\plotone{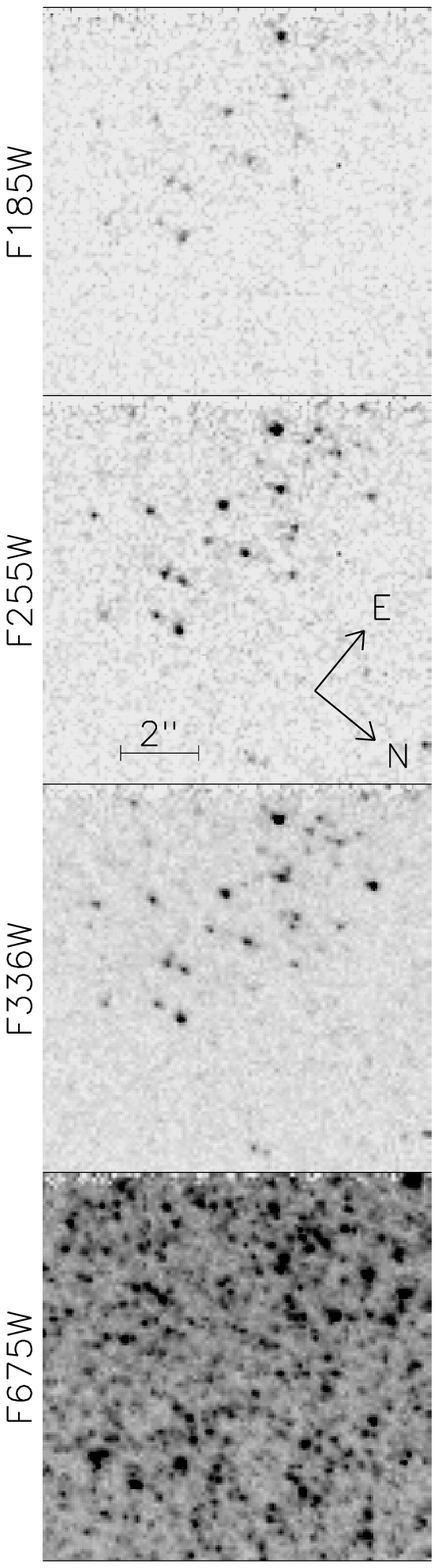}
\figcaption{WFPC2 images (10$''\times$10$''$ FOV) of the Archipelago region
(see text for details) in the F675W, F336W, F255W and F185W bands
respectively. The images have been resampled to a smaller pixel size as
described in Figure~\ref{mosaic_stars}.\label{mosaic_archipelago}}

\clearpage
\vspace*{3cm}
\epsscale{.7}
\plotone{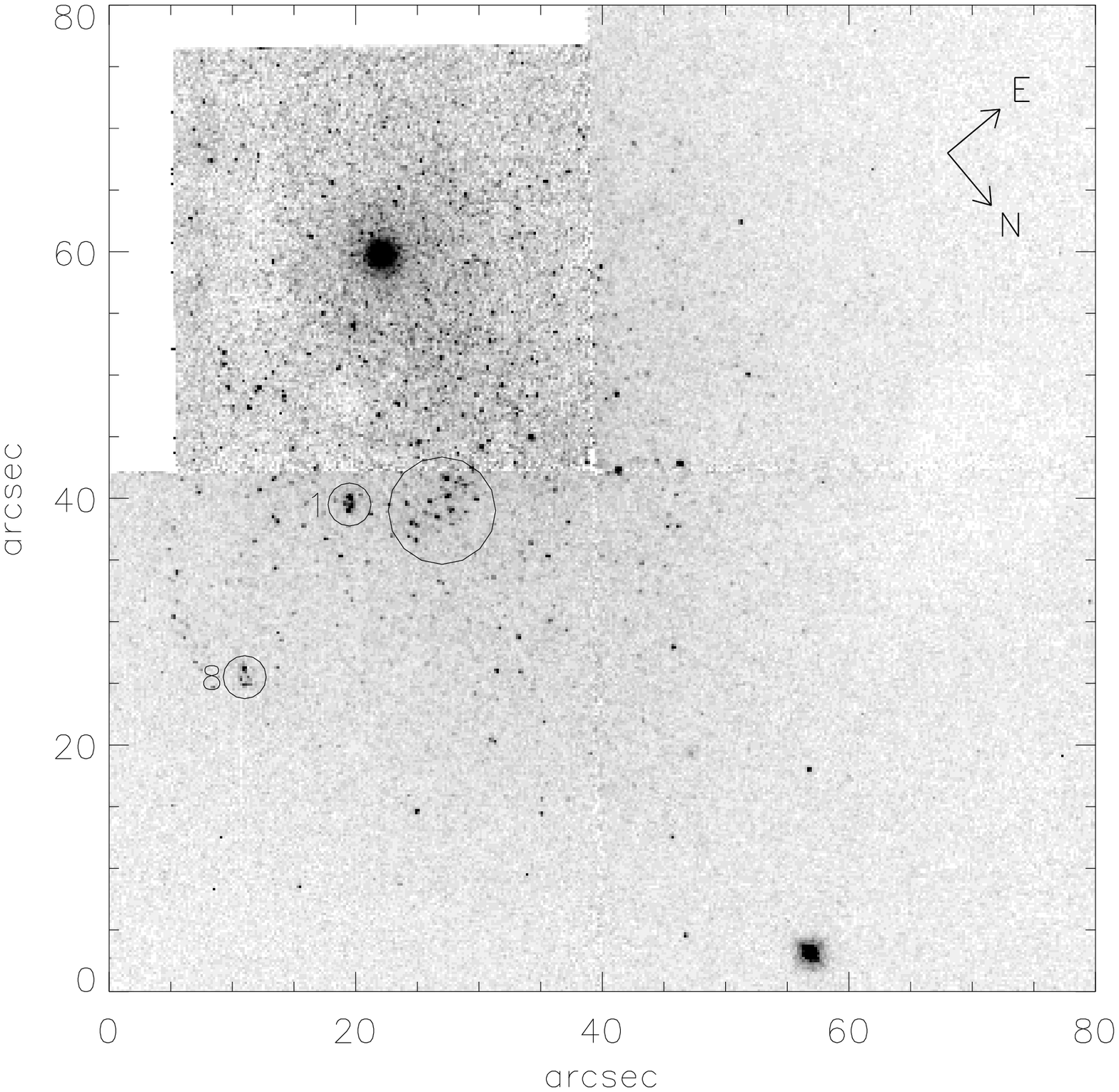}
\figcaption{An 80$''\times$80$''$ region taken from the full F336W WFPC2
mosaic. The nucleus of NGC~205 is visible at the center of the smaller PC1
CCD at the upper left corner of the picture. Clusters 1 and 8 are shown
inside the small circles. The larger circle marks the area of what we call
the ``Archipelago'' region (see text for details).\label{all_field}}

\clearpage
\vspace*{3cm}
\epsscale{.6}
\plotone{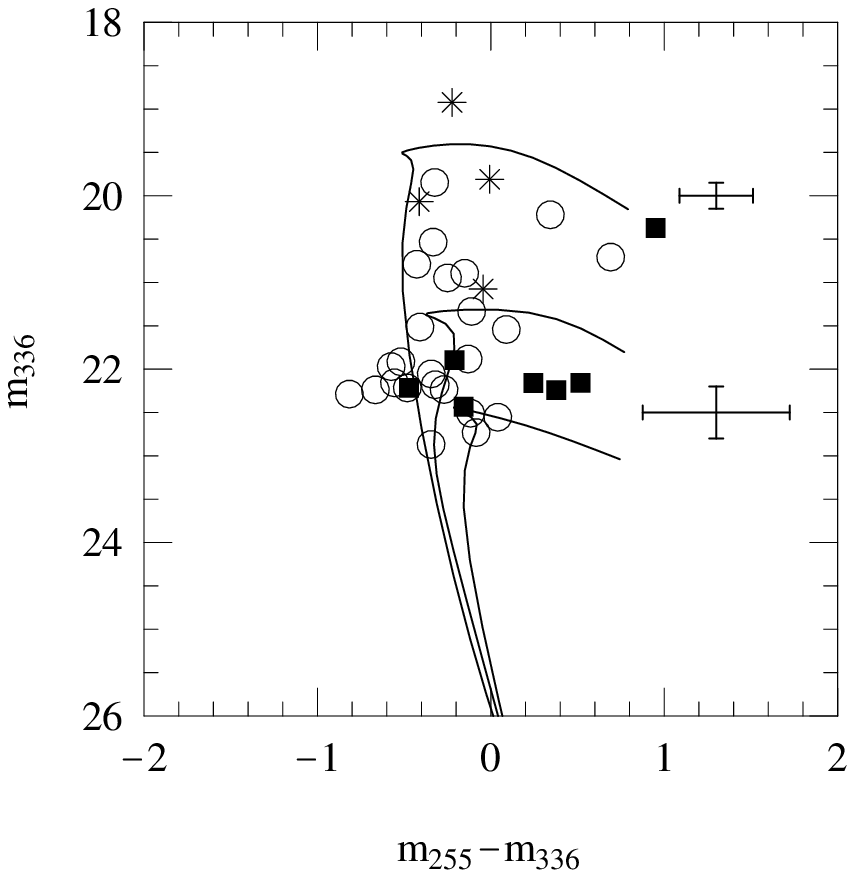}
\figcaption{The color-magnitude diagram ($m_{336}$ vs. $m_{255}-m_{336}$)
of Clusters 1 (asterisks) and 8 (filled squares), and of the Archipelago
(open circles). Note that only the four brightest stars of Cluster 1 have
been included in this diagram, as the photometry of the fainter stars
recognizable in Fig.~\ref{mosaic_stars} is affected by large errors. Mean
photometric errors are shown by the error bars to the right. Overplotted
are the theoretical isocrones from Bertelli et al.\ (1994) for the
metallicity $Z=0.02$ and ages of 20, 50 and 100 Myr respectively. These
isochrones have been tranformed into the STMAG magnitude system, by
adopting a true distance modulus of 24.6 mag and a reddening $E(\bv)=0.18$.
The post main-sequence part of the isocrones has been truncated at $T_{\rm
eff}=8,000$~K.\label{isocrone_n205}}

\end{document}